\documentclass[aps,pra,reprint,showkeys,amssymb,floatfix,longbibliography,superscriptaddress]{revtex4-2}

\usepackage{graphicx}
\usepackage{amsmath,amsfonts,amssymb}
\usepackage{hyperref}
\usepackage{braket}
\usepackage{xcolor}
\usepackage{tikz}
\usepackage{color}
\usepackage{enumerate}
\usepackage[caption=false]{subfig}
\usepackage{multirow}
\usepackage{qcircuit}
\usepackage[normalem]{ulem}
\usepackage{placeins}
\usepackage{url}
\usepackage{dsfont}
\usepackage{braket}
\usepackage[multidot]{grffile}
\usepackage{xspace}

\newcommand{\equref}[1]{Eq.~(\ref{#1})}

\newcommand\TWOOOQ{D-Wave 2000Q\xspace}
\newcommand\Advantage{Advantage\xspace}
\newcommand\const{\ensuremath{C}\xspace}

\begin{document}

\title{Benchmarking Advantage and D-Wave 2000Q quantum annealers with\\ exact cover problems}

\author{Dennis Willsch}
\thanks{Corresponding author: Dennis Willsch}
\email{d.willsch@fz-juelich.de}
\affiliation{Institute for Advanced Simulation, J\"ulich Supercomputing Centre,\\
Forschungszentrum J\"ulich, 52425 J\"ulich, Germany}
\author{Madita Willsch}
\affiliation{Institute for Advanced Simulation, J\"ulich Supercomputing Centre,\\
Forschungszentrum J\"ulich, 52425 J\"ulich, Germany}
\affiliation{AIDAS, 52425 J\"ulich, Germany}
\author{\firstname{Carlos D.} \surname{Gonzalez Calaza}}
\affiliation{Institute for Advanced Simulation, J\"ulich Supercomputing Centre,\\
Forschungszentrum J\"ulich, 52425 J\"ulich, Germany}
\author{Fengping Jin}
\affiliation{Institute for Advanced Simulation, J\"ulich Supercomputing Centre,\\
Forschungszentrum J\"ulich, 52425 J\"ulich, Germany}
\author{Hans De Raedt}
\affiliation{Institute for Advanced Simulation, J\"ulich Supercomputing Centre,\\
Forschungszentrum J\"ulich, 52425 J\"ulich, Germany}
\affiliation{Zernike Institute for Advanced Materials, University of Groningen, Nijenborgh 4, 9747 AG Groningen, The Netherlands}
\author{Marika Svensson}
\affiliation{Jeppesen, 411 03 Gothenburg, Sweden}
\affiliation{Department of Computer Science and Engineering, Chalmers University of Technology, SE-412 96 Gothenburg, Sweden}
\author{Kristel Michielsen}
\affiliation{Institute for Advanced Simulation, J\"ulich Supercomputing Centre,\\
Forschungszentrum J\"ulich, 52425 J\"ulich, Germany}
\affiliation{AIDAS, 52425 J\"ulich, Germany}
\affiliation{RWTH Aachen University, 52056 Aachen, Germany}

\date{\today}

\begin{abstract}
  We benchmark the quantum processing units of the
  largest quantum annealers to date, the 5000+ qubit quantum annealer \Advantage
  and its 2000+ qubit predecessor \TWOOOQ, using tail assignment and exact cover problems from aircraft scheduling scenarios. The benchmark set contains small, intermediate, and large problems with both sparsely connected and almost fully connected instances.
  We find that \Advantage outperforms \TWOOOQ for almost all problems, 
  with a notable increase in success rate and problem size. In particular, \Advantage is also able to solve the largest problems with 120 logical qubits that \TWOOOQ cannot solve anymore.
  Furthermore, problems that can still be solved by \TWOOOQ are solved faster by \Advantage.
  We find, however, that \TWOOOQ can achieve better success rates for sparsely connected problems that do not require the many new couplers present on \Advantage, so improving the connectivity of a quantum annealer does not per se improve its performance.
\end{abstract}


\maketitle

\section{Introduction}

Quantum annealing is a quantum computing paradigm that relies on quantum fluctuations to solve
optimization problems \cite{Apolloni89,finnila94,KadowakiNishimori1998QuantumAnnealing,Brooke99, Harris2010DWave,Johnson2011DWave, Bunyk2014DWave, Job2018TestDriving1000Qubits, Hauke2020PerspectivesQuantumAnnealing, Nath2021QMLForRealWorlApplications}.
In September 2020, D-Wave Systems has released a quantum annealer with a 5000+ qubit quantum processing unit (QPU)
called \Advantage \cite{dwave2020Advantage}. This system has more than twice as many qubits as its predecessor
\TWOOOQ and an increase in qubit
connectivity from 6 to 15 by using the Pegasus topology \cite{dwave2020AdvantageTechnologyUpdate, dwave2020TechnicalDescription,boothby2020nextgeneration,king2020performance}.
High expectations have been placed on its computational power, and first independent studies have become available
\cite{cohen2020picking,kuramata2020larger,calaza2021gardenoptimization, birdal2021QuantumSync, Bhatia2021PerformanceAnalysisQSVM,fox2021bio,rahman2021su2latticegaugetheorydwave,Phillipson2021QSVMAdvantageBenchmark}.
For such a rapidly developing technology, it is an important task for independent researchers to study progress and test new developments.

Conceptually, there are three classes of benchmarks for quantum annealers:
\begin{enumerate}
  \item[(1)] Comparison with detailed real-time simulations of quantum annealing systems based on solving the time-dependent Schr\"odinger equation \cite{Willsch2020FluxQubitsQuantumAnnealing, WillschMadita2020PhD} or the time-dependent master equation  \cite{Harris2010DWave,Johnson2011DWave,boixo13,Albash2015,albash15,boixo16,marshall19}.
  \item[(2)] Direct QPU benchmarks (including comparison with other quantum annealing systems and optimization problem solvers) for problems of intermediate size that may or may not need embeddings and solve either real-world or artificial problems \cite{boixo14,Ronnow14,Hall2015SpanningTreeDWave,hen15,McGeoch2015BenchmarkingDW2X, Novotny2016SpanninTreeDWave, Li2018TFDNABindingAffinity, Willsch2019BenchmarkingQAOA, Willsch2020QSVM, Cavallaro2020QSVMRemoteSensing, dwave2020Advantage,domino2021quantum}.
  \item[(3)] Benchmarks of hybrid solvers that use a combination of QPUs and CPUs or GPUs to solve large-scale application problems \cite{Mugel2020dynamicPortfolioOptimizationQAHybridTensorFinance, calaza2021gardenoptimization, dwave2020AdvantageTechnologyUpdate,grozea2021optimising}.
\end{enumerate}
The experiments reported in this paper focus on benchmarking the bare QPU performance, thus belonging to benchmarking class (2).

We assess the progress in quantum annealing technology by benchmarking both \Advantage and \TWOOOQ with exact cover problems. The exact cover problem is an NP-complete problem \cite{Karp1972KarpsNPCompleteProblems} that has become a prominent application to study quantum annealing 
\cite{Farhi2001QuantumAdiabaticNPComplete, choi2010adiabaticQuantumAlgorithms, Lucas2014IsingQUBOFormulationManyNPproblems, Cao2016SetCoverProblemsQA, Sax2020ApproximateApproximationQA}
and gate-based quantum computing 
\cite{Farhi2014QAOA, Vikstal2019QAOATailAssignment, Bengtsson2020QAOAExactCoverProblem, svensson2021LargeILPBranchPrice, Willsch2021JUQCSGQAOA}.
In our case, the exact cover problems represent simplified aircraft scheduling scenarios. They are derived from the tail assignment problem \cite{Groenkvist2005TailAssignmentProblem} (see \cite{svensson2021LargeILPBranchPrice} for more information), which belongs to the class of set covering and partitioning problems that is covered in a tremendous amount of literature on both applications and solution techniques in Operations Research (see e.g.~\cite{Ernst2004StaffSchedulingAndRostering,Tahir2019IntegralColumnGeneration}). Related aircraft assignment problems on quantum annealers have been studied in \cite{Stollenwerk2019FlightGateAssignmentQA, Stollenwerk2020QAAirTrafficManagement, Martins2020TailAssignmentProblemQA}.

\begin{figure*}
  \centering
  \includegraphics[width=2\columnwidth]{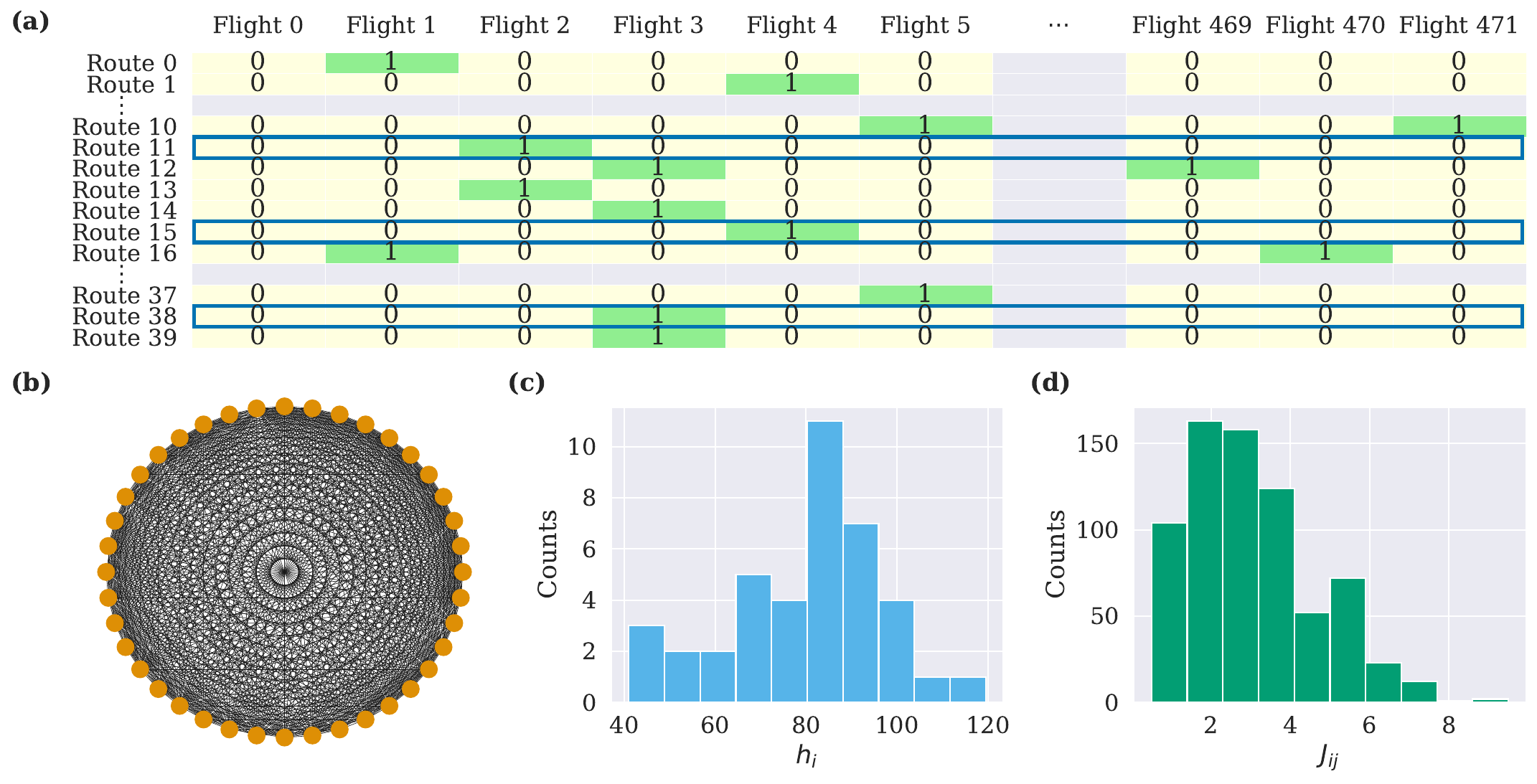}
  \caption{\textbf{Visualization of the exact cover problem with 40 logical qubits (instance 0).
  (a)} Boolean matrix $A$ defining the exact cover problem instance (see \equref{eq:exactcover}). For each of the 40 routes, the matrix $A$ indicates all flights that are covered by this route. The exact cover problem is to find a selection of routes (i.e., a subset of rows of the matrix $A$) such that all 472 flights are covered exactly once (meaning that the sum of the selected rows contains only ones). For this problem, rows belonging to this solution are indicated with blue boxes. The full solution is given by the ground state $\vert0100001010010101000000101000100000000000\rangle$ (the rightmost qubit corresponds to route 0).
  \textbf{(b)} Coupler graph of the Ising formulation of this problem (cf.~\equref{eq:Ising}), where each non-zero $J_{ij}$ corresponds to a black line between the 40 qubits.
  With 711 out of all $\binom{40}{2}=780$ couplers being non-zero, this problem is almost fully connected
  (see also Appendix~\ref{app:details}).
  A distribution of the values of the Ising parameters of this problem is shown in
  \textbf{(c)} for the qubit biases $h_i$ given by \equref{eq:TailAssignmentIsinghi} and
  \textbf{(d)} for the couplers $J_{ij}$ given by \equref{eq:TailAssignmentIsingJij}).}
  \label{fig:visualization}
\end{figure*}

The essence of the present exact cover problems is shown in Fig.~\ref{fig:visualization}(a).
In this case, we are given 40 flight routes. Each route contains several out of 472 flights.
The task is to find a selection of flight routes such that all 472 flights are
covered \emph{exactly} once.
On the quantum annealer, each route is represented by a qubit.
If a route is to be selected, the corresponding qubit ends up in the state $\ket 1$
after the measurement. Eventually, each selected route shall be assigned to one airplane. We remark that solving the problem with a given set of routes is a simplification of the general case. 

The difficulty of the problem for a quantum computer can be seen
by the following counting argument: For 40 routes (i.e.~40 qubits),
the number of possible selections is $2^{40}\approx10^{12}$.
For 120 routes (i.e.~120 qubits),
which are the largest problems that are solved in the present benchmark set,
the number of selections already grows to $10^{36}$.
Hence, exact cover problems are well suited for benchmarking the \Advantage and \TWOOOQ
QPUs.

We find that \Advantage outperforms \TWOOOQ on almost
all problems in the present benchmark set up to 120 logical qubits.
\Advantage can embed and solve larger problems. Furthermore,
the time-to-solution on \Advantage is at least roughly a factor of two shorter.
\Advantage scores better success rates for all problems with almost all-to-all connectivity.
Only some problems with a very sparse qubit connectivity have sometimes higher success
rates on \TWOOOQ.

The remainder of this paper is structured as follows. In Sec.~\ref{sec:methods}, we describe the
mathematical details associated with the exact cover problems under investigation.
In Sec.~\ref{sec:results}, we present and discuss the results that \Advantage and
\TWOOOQ have produced for both small-scale and large-scale
exact cover problems. Section~\ref{sec:conclusion} contains our conclusions.

\section{Methods}
\label{sec:methods}

This section presents the mathematical details behind the problems under investigation and how they are solved on the quantum annealers.
We first outline the type of optimization problems that \Advantage and \TWOOOQ can solve, including the important distinction between \emph{physical} and \emph{logical} qubits. Then, we describe the tail assignment and exact cover problems under investigation, along with their formulation on the \TWOOOQ and \Advantage QPUs.

\subsection{QUBO and Ising problems}
\label{sec:qa}

The QPUs produced by D-Wave Systems are designed to solve binary quadratic models (BQMs), i.e., quadratic optimization problems over discrete variables that can each take two different values. BQMs are typically formulated as quadratic unconstrained binary optimization (QUBO) models or Ising models:
\begin{align}
  \label{eq:QUBO}
  &\text{QUBO}: & &
\min_{x_i=0,1} \left(\sum_{i\le j} x_i Q_{ij} x_j + \const_1\right), \\
  \label{eq:Ising}
  &\text{Ising}: & &\min_{s_i=\pm1} \left(\sum_{i} h_i s_i + \sum_{i<j} J_{ij} s_i s_j + \const_2\right), 
\end{align}
where the indices $i$ and $j$ range over all qubits.
In the QUBO model in \equref{eq:QUBO}, the problem is defined by the QUBO matrix $Q$ with values $Q_{ij}\in\mathbb R$, and the binary problem variables are $x_i=0,1$. In the Ising model in \equref{eq:Ising}, the problem is defined by the biases $h_i\in\mathbb R$ and the couplers $J_{ij}\in\mathbb R$, and the problem variables are $s_i=\pm1$. An example distribution of $h_i$ and $J_{ij}$ of the problems under investigation is shown in Figs.~\ref{fig:visualization}(c) and (d), respectively.

It is worth mentioning that on the D-Wave QPUs, all problems are internally converted into Ising models and (if the autoscaling feature is on) rescaled by a constant factor such that $h_i\in[-2,2]$ and $J_{ij}\in[-1,1]$ \cite{DWaveSolversParameters} (note that these ranges might be different for future QPUs). To convert between QUBO
and Ising formulation, we use the quantum annealing convention (see also \cite{DWaveProblemSolvingHandbook})
\begin{align}
  \label{eq:QUBO2Ising}
  x_i = \frac{1+s_i} 2,
\end{align}
so that $x_i=0$ ($x_i=1$) maps to $s_i=-1$ ($s_i=1$). Note that in the literature on gate-based quantum computing, also the alternative $x_i=(1-s_i)/2$ is often used \cite{NielsenChuang} (which would result in a change of sign for the qubit biases, $h_i\mapsto-h_i$)

The constants $\const_1$ and $\const_2$ in Eqs.~(\ref{eq:QUBO}) and (\ref{eq:Ising}) do not affect the solution of the problem. However, they can be used to shift the energy (i.e., the value of the objective function at the solution). We use it to shift the energy of the ground state to zero so that we can conveniently determine the success rate by counting all solutions with energy zero.

\subsection{Physical and logical qubits}
\label{sec:qubits}

Many problems may require non-zero couplers $Q_{ij}$ or $J_{ij}$ between different qubits $i$ and $j$ that do not physically exist on the QPUs. For instance, the 40 qubit problem sketched in Fig.~\ref{fig:visualization}(b) has almost all-to-all connectivity. In this case, solving the problem on a QPU requires the concept of \emph{embedding} the problem on a QPU.

Conventionally, the qubits that physically exist on a QPU are called \emph{physical} qubits. On a \TWOOOQ QPU, the 2000+ physical qubits are connected in a \emph{Chimera} topology \cite{Bunyk2014DWave}. This means that each physical qubit is connected to 6 other physical qubits on average. On an \Advantage QPU, the Chimera topology has been upgraded to the \emph{Pegasus} topology \cite{dwave2020Advantage}. This means that nearly all of the 5000+ physical qubits are connected to 15 other physical qubits, increasing the connectivity by a factor of $2.5$.

When problems require a larger connectivity than provided by the QPU, the effective connectivity between the qubits can be increased by combining several physical qubits into a \emph{logical} qubit. The QUBO and Ising models in Eqs.~(\ref{eq:QUBO}) and (\ref{eq:Ising}) are typically formulated in terms of such logical qubits, and not the underlying physical qubits. The physical qubits that form a logical qubit are called a \emph{chain}. To ensure that physical qubits within a chain function as a single logical qubit, the couplers $J_{ij}$ between them are set to a reasonably large, negative value called $\texttt{chain\_strength}$.
If a chain between two qubits \emph{breaks} (i.e., if the product of the Ising variables is $s_is_j = -1$), a penalty of $2\times\texttt{chain\_strength}$ is added to the energy.

We define the $\texttt{chain\_strength}$ in terms of
the \emph{relative chain strength} $\mathrm{RCS\in[0,1]}$ according to
\begin{align}
  \label{eq:rcs}
  \texttt{chain\_strength} = \mathrm{RCS} \times
  \texttt{max\_strength},
\end{align}
where $\texttt{max\_strength}=\max\left(\left\{\left|h_i\right|\right\}\cup\left\{\left|J_{ij}\right|\right\}\right)$ is the maximum absolute value of all $h_i$ and $J_{ij}$. For instance, for the problem sketched in Fig.~\ref{fig:visualization},
\texttt{max\_strength} would be $119.5$.

An \emph{embedding} is a mapping from each logical qubit to a chain of physical qubits. An important property of embeddings are the chain lengths. In our experience, embeddings with too large chains may result in a poor quality of the solutions produced by a QPU. In Appendix~\ref{app:details}, we provide more details on the specific chains encountered in the embeddings (see Fig.~\ref{fig:problemdetails}).

The D-Wave Ocean SDK \cite{DWOceanSDK} provides algorithms to automatically generate qubit embeddings with a given value for $\texttt{chain\_strength}$. Still, when using a QPU, finding and characterizing embeddings is an important step and may considerably affect the quality of the solution. For this reason, as part of the present benchmark, we systematically investigate different embeddings and relative chain strengths in Sec.~\ref{sec:results3040}.

Note that the Pegasus topology is an extension of the Chimera topology, so that a Chimera graph can be natively embedded in a Pegasus graph (see Appendix~\ref{app:conjecture}  and also \cite{calaza2021gardenoptimization}). We made use of this relation for the experiments presented in Appendix~\ref{app:conjecture}. For all other experiments, the embeddings onto the Chimera and Pegasus topologies were generated independently using the D-Wave Ocean SDK.

\subsection{Tail assignment problem}
\label{sec:tailassignment}

The problem instances considered in this work are derived from the tail assignment problem \cite{Groenkvist2005TailAssignmentProblem}. The tail assignment problem is a fundamental component of aircraft assignment problems, i.e., the problem of assigning flights to individual airplanes, identified by their tail number.
The objective is to minimize the overall cost subject to certain constraints such as minimum connection times, airport curfews, maintenance, and preassigned activities. The general role of tail assignment problems in aircraft scheduling and their relation to the column generation technique \cite{Desrosiers2005ColumnGeneration} and the branch-and-price algorithm \cite{Barnhart1998BranchAndPrice} is described in detail in \cite{svensson2021LargeILPBranchPrice}.   

We consider a simple form of the tail assignment problem given by
\begin{align}
  \label{eq:TailAssignmentProblemCost}
  &\text{minimize} & &\sum_{r=0}^{R-1} c_r x_r, & \\
  \label{eq:TailAssignmentProblemConstraints}
  &\text{subject to} & &\sum_{r=0}^{R-1} A_{rf} x_r = 1 \quad \forall f=0,\ldots,F-1, &
\end{align}
where $r=0,\ldots,R-1$ enumerates all flight routes, $f=0,\ldots,F-1$ enumerates the flights, $x_r\in\{0,1\}$ are the Boolean problem variables with $x_r=1$ if route $r$ is to be selected, $c_r$ is the cost of selecting route $r$, and $A\in\{0,1\}^{R\times F}$ is the Boolean problem matrix with $A_{rf}=1$ if flight $f$ is contained in route $r$
(see Fig.~\ref{fig:visualization}(a) for an example of the matrix $A$).
Further models for the tail assignment problem can be found in \cite{Groenkvist2005TailAssignmentProblem,Vikstal2019QAOATailAssignment,Martins2020TailAssignmentProblemQA,svensson2021LargeILPBranchPrice}.

A BQM version of the tail assignment problem given by Eqs.~(\ref{eq:TailAssignmentProblemCost}) and (\ref{eq:TailAssignmentProblemConstraints}) is
\begin{align}
  \label{eq:TailAssignmentBQM}
  &\min_{\vec x\in \{0,1\}^N}\left(\lambda\vec c^T\vec x + \left(A^T\vec x - \vec b\right)^2\right),
\end{align}
where the number of qubits $N=R$ is given by the number of flight routes,
$\vec c = (c_0,\ldots,c_{R-1})^T$ contains the costs, $\vec b = (1,\ldots,1)^T$ is an $F$-dimensional vector of ones.
Note that the scaling factor $\lambda$ in \equref{eq:TailAssignmentBQM} is the inverse of the penalty multiplier that would determine the scale of the constraint $A\vec x=\vec b$ in \equref{eq:TailAssignmentProblemConstraints}. It was put in front of the objective function $\vec c^T\vec x$ so that the exact cover version of the problem corresponds to $\lambda=0$ (see Sec.~\ref{sec:ec}).

We obtain the QUBO formulation of the tail assignment problem by multiplying out the square in \equref{eq:TailAssignmentBQM} and collecting all terms into the general QUBO model \equref{eq:QUBO}. An outline of the calculation is given in Appendix~\ref{app:derivation}. After doing this, we can read off the entries of the QUBO matrix as
\begin{align}
  \label{eq:TailAssignmentQUBOQij}
  Q_{ij} &= \begin{cases}
  (2AA^T)_{ij} & (i<j) \\
  (AA^T)_{ii} - (2A\vec b)_i + \lambda c_i & (i=j)
  \end{cases},\\
  \label{eq:TailAssignmentQUBOC1}
  \const_1 &= \vec b^T\vec b.
\end{align}
Note that the qubit indices $i,j\in\{0,\ldots,N-1\}$ correspond to the previous route index $r\in\{0,\ldots,R-1\}$.

Finally, the Ising formulation of the tail assignment problem is found by using \equref{eq:QUBO2Ising} to replace the qubit variables $x_i$ by spin variables $s_i$ in the QUBO model \equref{eq:QUBO}. After collecting linear, quadratic, and constant terms, we find the expressions for the coefficients of the Ising model \equref{eq:Ising},
\begin{align}
  \label{eq:TailAssignmentIsinghi}
  h_i &= \sum_j \frac 1 2(AA^T)_{ij} - (A\vec b)_i + \frac 1 2 \lambda c_i, \\
  \label{eq:TailAssignmentIsingJij}
  J_{ij} &= \frac 1 2(AA^T)_{ij},  \\
  \label{eq:TailAssignmentIsingC2}
  \const_2 &= \const_1 + \sum_{i< j} \frac{1}2 (AA^T)_{ij} \nonumber\\
  &+ \sum_i \frac{1}2 ((AA^T)_{ii} - (2A\vec b)_i + \lambda c_i).
\end{align}
These expressions hold for general values of $A$ and $\vec b$ (details on the calculation can also be found in Appendix~\ref{app:derivation}).
A characteristic distribution of the biases $h_i$ and the couplers $J_{ij}$ is shown in Figs.~\ref{fig:visualization}(c) and (d), respectively.

\subsection{Exact cover problem}
\label{sec:ec}

The exact cover problem is an NP-complete set partitioning problem  \cite{Karp1972KarpsNPCompleteProblems}. In matrix form, it can be written as
\begin{align}
  \label{eq:exactcover}
   \min_{x_r=0,1} \sum_{f=0}^{F-1} \left( \sum_{r=0}^{R-1}A_{rf}x_r - 1 \right)^2,
\end{align}
and its purpose can be directly understood from Fig.~\ref{fig:visualization}(a):
The selection of routes (i.e., rows) with $x_r=1$ has to be such that each flight $f$ in the problem matrix $A$ is covered exactly once.

The exact cover problem corresponds to the feasibility version of the tail assignment problem given by the sole constraints in \equref{eq:TailAssignmentProblemConstraints}, without the objective function in \equref{eq:TailAssignmentProblemCost}.
Formally, the exact cover problems are obtained by setting $\lambda=0$ in \equref{eq:TailAssignmentBQM}, which yields \equref{eq:exactcover}.

Hence, the QUBO (Ising) coefficients of the exact cover problem are given by \equref{eq:TailAssignmentQUBOQij} (Eqs.~(\ref{eq:TailAssignmentIsinghi}) and (\ref{eq:TailAssignmentIsingJij})) for $\lambda=0$ (see also \cite{Lucas2014IsingQUBOFormulationManyNPproblems}).
Note that the explicit expressions for the constants in Eqs.~(\ref{eq:TailAssignmentQUBOC1}) and (\ref{eq:TailAssignmentIsingC2}) are not relevant for the solution of the problem on the quantum annealer. However, it is convenient to add them to the resulting energies to ensure that the energy minimum is zero, because then we can determine the success rate by counting the occurrences of samples with energy zero.

The problems have been generated using the Column Generation method as described in detail in \cite{svensson2021LargeILPBranchPrice}. Additional properties of the exact cover problems, including the number of logical couplers and physical qubits used in the generated embeddings, can be found in Appendix~\ref{app:details}.

\section{Results}
\label{sec:results}

\begin{figure}
  \centering
  \includegraphics[width=\columnwidth]{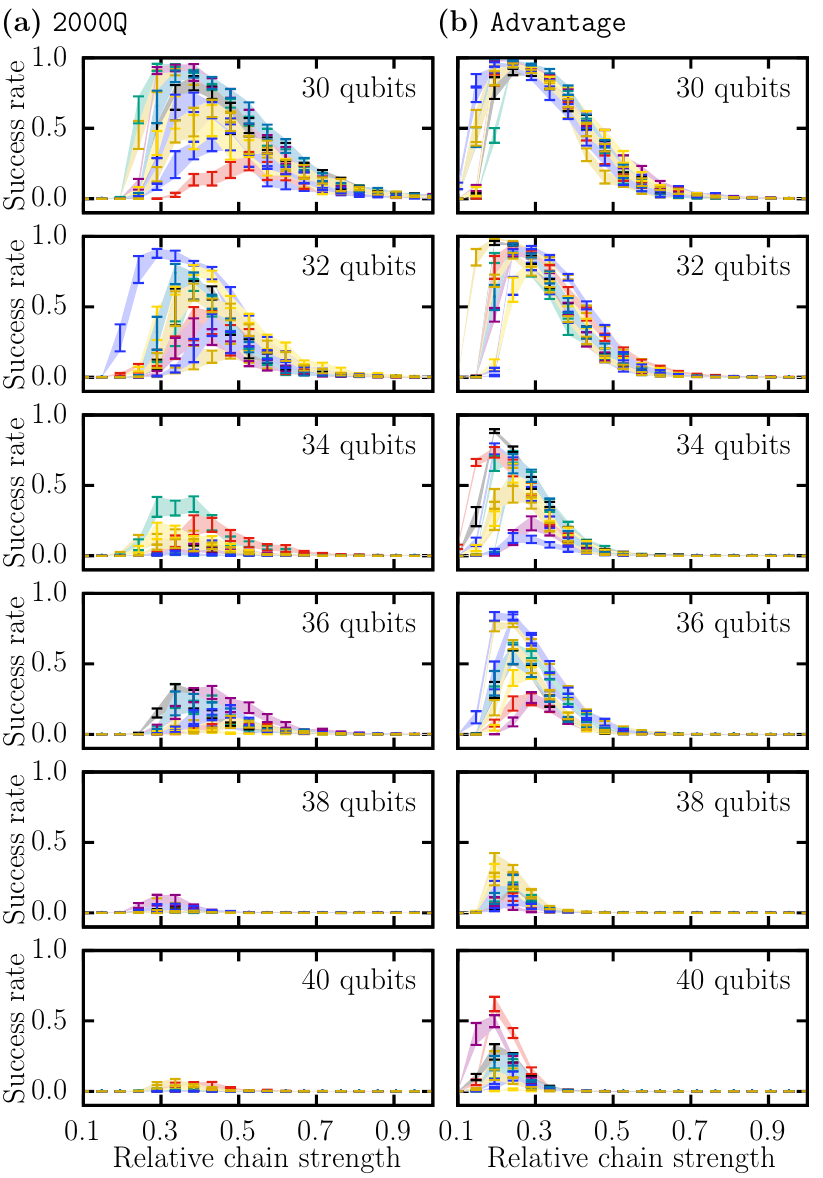}
  \caption{\textbf{Success rates for exact cover problems with 30--40 qubits (instance 0) as a function of the relative chain strength (bottom axis) on (a) \TWOOOQ and (b) \Advantage with default annealing time $20\,\mathrm{\mu s}$.} The scan of the relative chain strength is repeated for 10 different, randomly generated embeddings (represented by different colors) and with 10 repetitions each to gather statistics. Markers indicate the corresponding standard deviation above and below the mean. Filled areas between the markers are guides to the eye. The curves for the success rates as a function of the RCS are representative of the other problem instances 1, 2, and 3 characterized in Appendix~\ref{app:details}.}
  \label{fig:scanrcs}
\end{figure}
\begin{figure}
  \centering
  \includegraphics[width=\columnwidth]{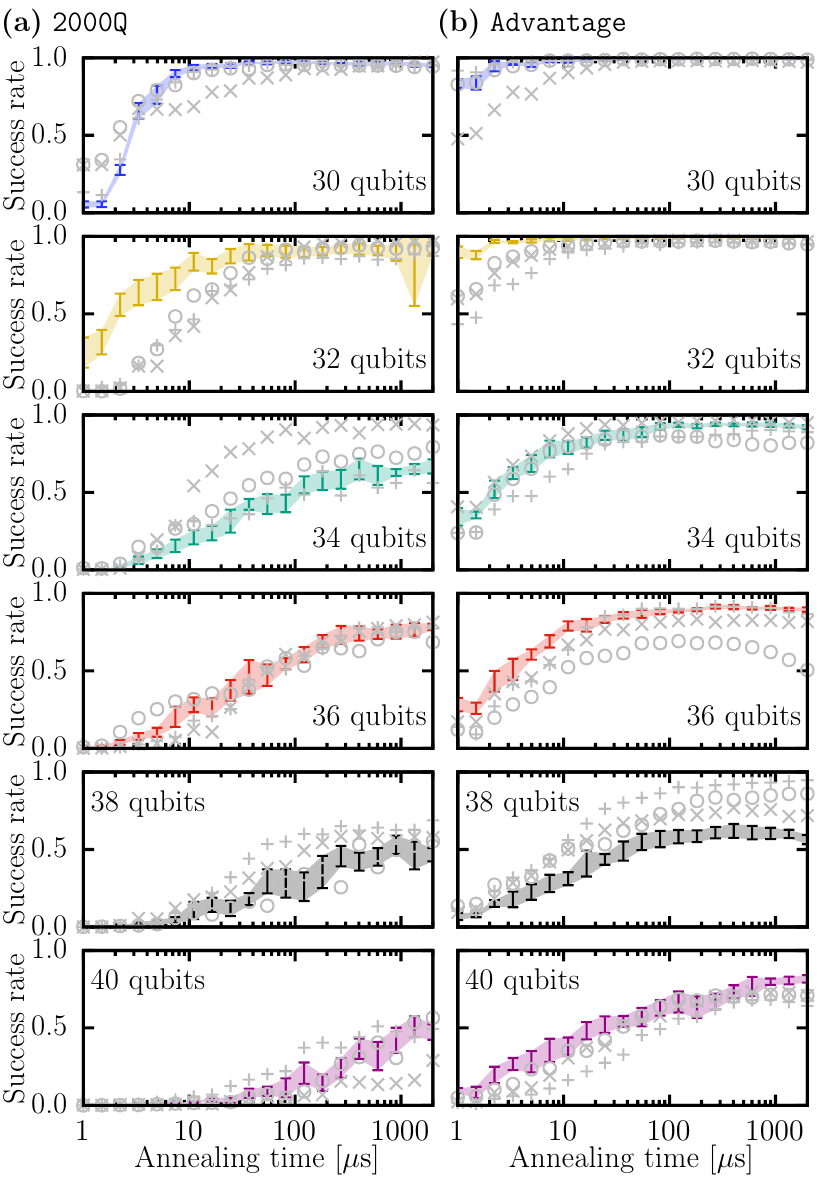}
  \caption{\textbf{Success rates as a function of the annealing time on (a) \TWOOOQ and (b) \Advantage, averaged over 10 repetitions.} Different markers indicate different problem instances: 0 (markers with error bars), 1 (pluses), 2 (crosses), 3 (circles). For each instance, the runs in each panel are performed with the same embeddings and relative chain strengths, characterized in Fig.~\ref{fig:problemdetails} in App.~\ref{app:details}. For instance 0 (highlighted in color), these parameters correspond to the best configuration of the corresponding panel in Fig.~\ref{fig:scanrcs}. Additionally, for instance 0, the standard deviation from the 10 repetitions is indicated by the filled areas between the error bars.}
  \label{fig:scanat}
\end{figure}

In this section, we present the benchmark results for \Advantage and \TWOOOQ. We first consider small and intermediate exact cover problems with 30--40 logical qubits and almost full connectivity, with a focus on comparing different embeddings and annealing times. Then, we proceed to large exact cover problems with up to 120 logical qubits, with a focus on the success rate and the time that it takes the QPUs to solve the problems.
We note that all considered problem sizes, although large for current quantum annealers, are still small for classical solvers and can be solved within a few minutes using for instance a linear programming solver.
For the sake of completeness, we also present results for tail assignment problems with $\lambda\neq0$ (cf.~\equref{eq:TailAssignmentBQM}) in Appendix~\ref{app:lambda}.

The experiments reported in this section were performed in August and September 2020 using the solver \texttt{DW\_2000Q\_VFYC\_6} for \TWOOOQ and the solvers \texttt{Advantage\_beta} and \texttt{Advantage\_system1.1} for \Advantage.
We stress that the goal of the present study was to compare the different QPUs under equivalent conditions. For this reason, we did not make use of any post-processing features to solely improve the quality of the results. Unless otherwise stated, all QPU settings were left at their default values.

\subsection{Densely connected problems with 30--40 qubits}
\label{sec:results3040}

The problems with $N=30,32,34,36,38,40$ qubits and almost full connectivity (cf.~Appendix~\ref{app:details}) are exact cover problems from aircraft scheduling
scenarios with 472 flights. Each qubit represents a flight route that contains some
of the 472 flights (cf.~Fig.~\ref{fig:visualization}). 
We remark that by construction, the ground state of each problem instance is unique and contains 9 qubits in state $\ket 1$.
We obtain the success rate by counting the number of samples with energy zero (note that the value of the constant \equref{eq:TailAssignmentIsingC2} can be used to shift the energy accordingly).

\subsubsection{Characterizing embeddings and chain strengths}

For each problem with $N=30,32,34,36,38,40$ qubits, we generate 10 different embeddings on both \TWOOOQ and \Advantage. For each embedding, we scan the RCS (cf.~\equref{eq:rcs}). The results are shown in Fig.~\ref{fig:scanrcs}.

We see that the results on \Advantage (Fig.~\ref{fig:scanrcs}(b)) are generally much better than on \TWOOOQ, especially for larger problems. The reason for this is that the 30--40 qubit problems are almost fully connected (cf.~Figs.~\ref{fig:visualization}(b) and \ref{fig:nonzerocouplers}). Hence, in such cases, the user can clearly profit from the much larger connectivity between qubits on the Pegasus topology.

This observation is in line with the fact that the chains on \Advantage for the same problem are much shorter (see Fig.~\ref{fig:problemdetails}(b)). We also see that for increasing problem size, generating multiple embeddings and tuning the chain strengths is crucial to obtain good results when using the bare QPUs.

\subsubsection{Varying the annealing time}

Next, we study the influence of the annealing time on the success rate. The reason is that for a quantum annealer, the annealing time is a central parameter with an expectable influence on the success rate: from the adiabatic theorem \cite{born1928adiabatictheorem}, comparably long annealing times are expected to lead to reasonable success rates. Given that the system is not completely isolated from an environment, however, long annealing times may lead to an increasing influence of noise on the system, and thus in fact degrade the performance. Therefore, experiments beyond the default annealing time are vital for reliable benchmarks of quantum annealers.

We first select the best embedding and relative chain strength for each problem, based on the results from the previous section. For problem instance 0, for example, this configuration corresponds to the position of the peaks in each panel in Fig.~\ref{fig:scanrcs}. For the selected configuration, we then replace the default annealing time of $20\,\mathrm{\mu s}$ by 20 different, logarithmically spaced annealing times in the QPU annealing time range $[1\,\mathrm{\mu s}, 2000\,\mathrm{\mu s}]$. The results for the 30--40 qubit problems and all four problem instances are shown in Fig.~\ref{fig:scanat}.

Comparing \TWOOOQ and \Advantage, we can make the following observations: First, \Advantage reaches higher maximum success rates, typically for the longest annealing time. Second, the success rates on \Advantage are already reasonable for short annealing times. This means that \Advantage is typically faster than \TWOOOQ (see also the comparison of QPU access times in Fig.~\ref{fig:dwavelargeproblems}(b) the following section). Finally, the fluctuations over 10 repetitions are smaller on \Advantage (see the filled areas in Fig.~\ref{fig:scanat}). We note that this observation also holds for the three instances whose statistics are not indicated in Fig.~\ref{fig:scanat} (gray markers). Thus, we conclude that \Advantage shows a demonstrable advantage over \TWOOOQ.

\subsection{Large problems with 50--120 qubits}

In this section, we consider exact cover problems of a larger size with 50 to 120 logical qubits. The goal is to assess
the performance of the QPUs for larger but more sparsely connected problem instances.
For each problem size, we consider six problem instances. Each corresponds to an
aircraft scheduling problem of the type sketched in Fig.~\ref{fig:visualization}(a)
with 535 flights. As before, the ground state of these exact cover problems is unique and known. It has 40 qubits in state $\ket 1$.

These large problems require many more physical qubits so that their embedded versions can occupy a large part of the QPUs. For this reason, the success rates may be smaller, and especially \TWOOOQ may not be able to solve the largest problems anymore.

\begin{figure}
  \centering
  \includegraphics[width=\columnwidth]{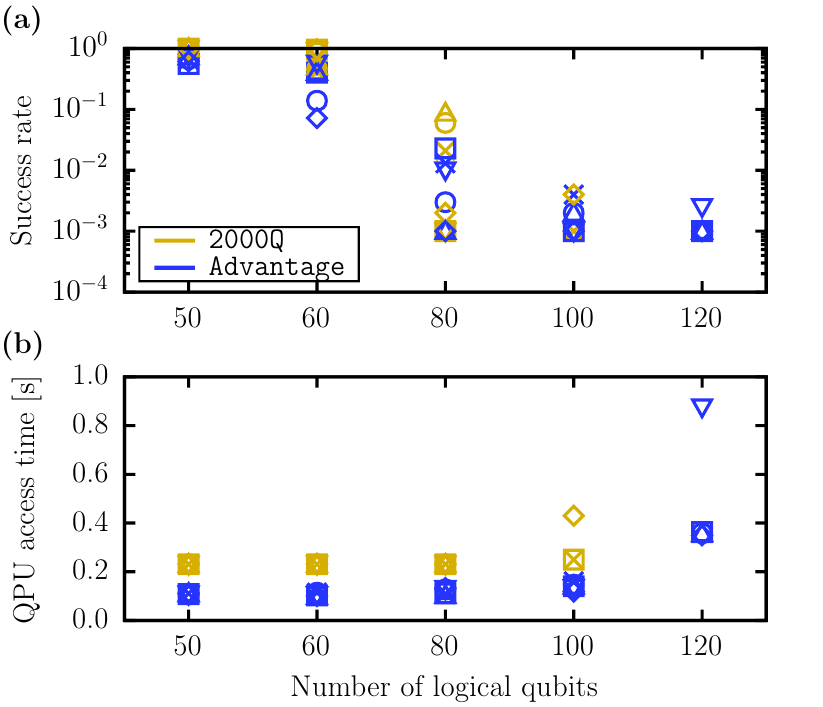}
  \caption{\textbf{Solution of exact cover problems with 50--120 logical qubits on both \TWOOOQ (yellow) and \Advantage (blue).}
  Shown are \textbf{(a)} the success rate as a function of the number of logical qubits and \textbf{(b)} the QPU access time (which mainly consists of programming, annealing, and readout times \cite{DWaveSolverComputationTime}) .
  Different markers indicate different problem instances:
  0 (crosses), 1 (squares), 2 (circles), 3 (up-pointing triangles), 4 (down-pointing triangles), 5 (diamonds).
  The embedding and the relative chain strength for these runs was selected using the same procedure as in Sec.~\ref{sec:results3040}.}
  \label{fig:dwavelargeproblems}
\end{figure}

Indeed, as Fig.~\ref{fig:dwavelargeproblems}(a) shows, only \Advantage
can solve five out of the six 120 qubit instances. The success rates for the 80 and
100 qubit instances are comparable for \TWOOOQ and \Advantage.
Only for the 50 and 60 qubit instances, we see that \TWOOOQ yields sometimes
better success rates than \Advantage.

The reason for this can be understood by studying the number of couplers required for these
problems: As shown in Fig.~\ref{fig:nonzerocouplers} in Appendix~\ref{app:details}, the 50 and 60 qubit
problems need comparably little couplers, so they have a sparser connectivity. Therefore, many of the physical couplers present in the
Pegasus topology on \Advantage are not required. In other words, the problems can already be embedded well enough on the Chimera topology.  This observation is
in line with a similar observation for \Advantage reported in \cite{calaza2021gardenoptimization}.

We conjecture that in this case, the additional
unused connections between qubits on \Advantage may disturb the annealing process (additional experimental evidence for this conjecture is given in Appendix~\ref{app:conjecture}), because even if they are not used, they still exist physically. The reason for our conjecture lies in a well-known physical mechanism: If two subsystems of a larger system are coupled, the system's Hamiltonian contains a corresponding interaction term. For flux qubits manufactured by D-Wave Systems, this interaction is often mediated by tunable interqubit inductive couplers \cite{Harris2010DWave}. Such an interaction is always on, due to the existence of a physical link between the subsystems, even if its scale is very weak or if the coupler is ``programmed to zero''. During the time evolution (which may take comparably long for an annealing process), the wave function of the system typically develops an error over time, which may have a detrimental impact on the function of a quantum computer. The effect can be easily observed for a system of coupled spins, for example the NMR quantum computing system \cite{DeRaedt2002QuantumSpinDynamics}. Furthermore, it can also be found in superconducting qubit systems; see e.g.~the crosstalk experiment in \cite{WillschDennis2020Phd}, the performance of two-qubit gates mediated by a coupler such as the resonator in \cite{Willsch2017GateErrorAnalysis} or the flux-tunable transmon in \cite{Lagemann2021NumericalAnalysisFluxTunbale}, or the theoretical study on $ZZ$ couplings in \cite{Xu2021ZZFreedomInTwoQubitGates}. We leave a detailed simulation of the present system for future research.

Finally, the large problems may require more time on the QPUs, so
that they are well suited to compare QPU run times between \TWOOOQ
and \Advantage.
Unless mentioned otherwise, the same number of reads and annealing times have been used for \TWOOOQ and \Advantage to ensure a fair comparison. The timing results are shown in Fig.~\ref{fig:dwavelargeproblems}(b).

For 50--80 qubits, the annealing time is less than or equal to $20\,\mathrm{\mu s}$,
so for 1000 reads it does not make up a significant fraction of the QPU access time.
Under these conditions, we see that \TWOOOQ still needs at least twice as long to solve the
problem. Thus, we infer that it is a speedup in programming and readout times \cite{DWaveSolverComputationTime} that make \Advantage faster than \TWOOOQ.

For problems with 100 and 120, the annealing time needs to be significantly increased to find the
ground state, which is visible in the QPU access times. For instance, problem instance 5 for 100 qubits on \TWOOOQ
(the single yellow diamond at 100 qubits in Fig.~\ref{fig:dwavelargeproblems}(b)) corresponds to
an annealing time of $200\,\mathrm{\mu s}$. The same annealing time is required for
problem instances 1, 2, 3, and 5 for 120 qubits on \Advantage (the blue cluster at 120 qubits in Fig.~\ref{fig:dwavelargeproblems}(b)).
Still, \Advantage solves these problems faster than \TWOOOQ solves the
corresponding 100 qubit problem. Only instance 4 for 120 qubits stands out (the blue
down-pointing triangle in Fig.~\ref{fig:dwavelargeproblems}(b)): Here, an
annealing time of $2000\,\mathrm{\mu s}$ was required to find a solution. In order not to
exceed the maximum run time on \Advantage, the number of reads was
reduced to 400. Thus, $400\times2000\,\mathrm{\mu s}=0.8\,\mathrm s$ makes up the
largest part of the QPU access time in this case.

\section{Conclusion}
\label{sec:conclusion}

In this paper, we have benchmarked the performance of the 2000+ qubit quantum annealer \TWOOOQ and the 5000+ qubit quantum annealer \Advantage. The benchmark suite consists of intermediate and large exact cover problems from  aircraft scheduling scenarios with both sparse and dense logical qubit connectivity.

We observed a considerable increase in performance on \Advantage. First, \Advantage was able to solve exact cover problems with up to 120 logical qubits that were unsolvable on \TWOOOQ. Second, the success rates produced by \Advantage were almost always higher. Third, \Advantage is approximately twice as fast as \TWOOOQ in terms of both programming and readout times. Additionally, the required annealing times to solve a problem on \Advantage are often shorter. Finally, the fluctuations in success rates over several repetitions on \Advantage were smaller.

A large part of the increase in performance can be attributed not only to the larger number of physical qubits, but rather to the increase in qubit connectivity: Every qubit in the Pegasus topology is connected to 15 other qubits, as compared to 6 other qubits in the Chimera topology used in \TWOOOQ.
We observed chain lengths in the embeddings that were roughly smaller by a factor of two.
We could only observe better performance on \TWOOOQ for problems with very sparse qubit connectivity or when using the same embedding on \TWOOOQ and \Advantage. The sparse problems could already be well embedded on the Chimera topology. We conclude that increasing the number of couplers does not necessarily improve the performance of a quantum annealer. Instead, whether an improvement or a degradation is observed depends on the connectivity of each problem instance. On the one hand, increasing the number of couplers allows to embed bigger problem instances, and it may also reduce the required chain length and thus the chain strength. On the other hand, we observed that if only a few couplers are needed, it may be better to use a processor with lower connectivity (see also \cite{calaza2021gardenoptimization}).

We also observed that, when the same embedding is used on the \TWOOOQ processor with connectivity 6 and the \Advantage processor with connectivity 15, the results on the \TWOOOQ processor are significantly better (see Appendix~\ref{app:conjecture}), suggesting that the additional couplers introduce additional noise if they are not needed.
In particular, this supports the conclusion that for our experiments, the better embeddings (and not a reduced noise level) are the reason for the improved performance of \Advantage over \TWOOOQ. Thus, a higher connectivity only yields an advantage if the embedding of the problem can actually be improved. An interesting project for future work would be a systematic study of the question whether, for a given embedding, there is some ``minimal improvement'' required to overcome the influence of additional unused couplers to also yield an improvement in the success probability.

When using the bare QPUs, it is essential to scan several embeddings and chain strengths to find optimal results. Furthermore, it is important to tune the annealing time. We note that besides the bare QPUs, we have also submitted all exact cover problems
of our benchmark set to D-Wave's hybrid solver services, which use a combination of QPUs and classical solvers to solve much larger problems \cite{dwave2020AdvantageTechnologyUpdate}.
All exact cover problems
could be solved by the hybrid solvers
\texttt{hybrid\_v1} (using \TWOOOQ) and
\texttt{hybrid\_binary\_quadratic\_model\_version2} (using \Advantage)
on September 14, 2020. See \cite{calaza2021gardenoptimization} for more detailed benchmarks of the hybrid solvers with problems of up to 12000 variables.

Our benchmark study confirms the consistent increase in both size and performance of quantum annealers over the past years. For the future, it is an interesting question whether D-Wave Systems will prove capable of keeping up the steep progress of doubling qubit numbers and increasing performance and qubit connectivity at the same time.

\section*{Data availability}

The datasets generated during and/or analyzed during the current study are available from the corresponding author on reasonable request.

\begin{acknowledgments}
We thank Manpreet Jattana for presenting parts of the results reported in this paper on Qubits 2020.
We thank D-Wave Systems for early access to an Advantage\textsuperscript{TM} system and the new hybrid solver service in Leap during the Advantage beta program. 
We gratefully acknowledge the J\"ulich Supercomputing Centre  for funding this project by providing additional computing time through the J\"ulich UNified Infrastructure for Quantum computing (JUNIQ) on the D-Wave quantum annealer.
D.W. and M.W. acknowledge support from the project JUNIQ that has received funding from the German Federal Ministry of Education and Research (BMBF) and the Ministry of Culture and Science of the State of North Rhine-Westphalia.
C.G. acknowledges support from the project OpenSuperQ (820363) of the European Quantum Flagship.
M.S. acknowledges funding from the Knut and Alice Wallenberg foundation (KAW) through the Wallenberg Centre for Quantum Technology (WACQT).
\end{acknowledgments}

\bibliographystyle{apsrev4-2custom}
\bibliography{bibliography}
\onecolumngrid
\appendix

\section{Derivation of QUBO and Ising models}
\label{app:derivation}

In this appendix, we outline the derivation of the QUBO and Ising coefficients from the BQM in \equref{eq:TailAssignmentBQM} for the tail assignment and exact cover problems under investigation.

To obtain the QUBO formulation, we first multiply out the square in \equref{eq:TailAssignmentBQM},
\begin{align} 
    \left(A^T\vec x - \vec b\right)^2 
    &= \left(A^T\vec x - \vec b\right)^T\left(A^T\vec x - \vec b\right) \\
    &= \vec x^T AA^T\vec x - \vec b^TA^T\vec x - \vec x^T A \vec b + \vec b^T\vec b \\
    \label{eq:appqubo}
    &= \sum_{ij} x_i (AA^T)_{ij} x_j - \sum_i (2A\vec b)_i x_i + \vec b^T\vec b.
\end{align}
After splitting the first sum into three parts, $\sum_{ij} = \sum_{i<j}+\sum_{i>j}+\sum_i$, and exchanging $i\leftrightarrow j$ in the second part (since the matrix $AA^T$ is symmetric), we obtain the upper triangular coefficients of the QUBO matrix, $Q_{ij} = (2AA^T)_{ij}$ for $i<j$, and the first part of the diagonal coefficients $(AA^T)_{ii}$. The second sum in \equref{eq:appqubo} yields, after using $x_i=x_ix_i$, the remaining part of the diagonal coefficients. The last term yields the constant contribution to the QUBO model. Combining this with the linear term $\lambda \vec c^T\vec x$ in \equref{eq:TailAssignmentBQM}, we obtain all coefficients of the QUBO model $\sum_{i\le j} x_i Q_{ij} x_j + \const_1$,
\begin{align}
  Q_{ij} &= \begin{cases}
  (2AA^T)_{ij} & (i<j) \\
  (AA^T)_{ii} - (2A\vec b)_i + \lambda c_i & (i=j)
  \end{cases},\\
  \const_1 &= \vec b^T\vec b = F.
\end{align}

To obtain the coefficients $h_i$ and $J_{ij}$ of the corresponding Ising model, we replace the qubit variables by spin variables, $x_i = (1+s_i)/2$ (cf.~\equref{eq:QUBO2Ising}),
\begin{align}
    \sum_{i\le j} x_i Q_{ij} x_j &= \sum_{i\le j} \frac{(1+s_i)}2 Q_{ij} \frac{(1+s_j)}2 \\
    &= \sum_{i< j} \frac{(1+s_i)}2 (2AA^T)_{ij} \frac{(1+s_j)}2 
    + \sum_i \frac{(1+s_i)}2 ((AA^T)_{ii} - (2A\vec b)_i + \lambda c_i) \frac{(1+s_i)}2\\
    &= 
    \sum_{i< j} \frac{1}2 (AA^T)_{ij}s_i + \sum_{i> j} \frac{1}2 (AA^T)_{ij} s_i
    + \sum_i \frac{1}2 (AA^T)_{ii}s_i - \sum_i (A\vec b)_i s_i + \sum_i \frac{1}2 \lambda c_i s_i\\
    &+ \sum_{i< j} \frac 1 2 (AA^T)_{ij}s_is_j\\
    &+ \sum_{i< j} \frac{1}2 (AA^T)_{ij}
    + \sum_i \frac{1}2 ((AA^T)_{ii} - (2A\vec b)_i + \lambda c_i),
\end{align}
where we used that the matrix $AA^T$ is symmetric and that $s_i^2=1$. Note that this calculation does not make use of prior knowledge about the values of $A$ and $\vec b$.

We can then identify the coefficients of the Ising model $\sum_{i} h_i s_i + \sum_{i<j} J_{ij} s_i s_j + \const_2$ as
\begin{align}  
  h_i &= \sum_j \frac 1 2(AA^T)_{ij} - (A\vec b)_i + \frac 1 2 \lambda c_i, \\
  J_{ij} &= \frac 1 2(AA^T)_{ij}, \\
  \const_2 &= \const_1 + \sum_{i< j} \frac{1}2 (AA^T)_{ij}
    + \sum_i \frac{1}2 ((AA^T)_{ii} - (2A\vec b)_i + \lambda c_i).
\end{align}

\section{Additional experiments for the conjecture about unused couplers}
\label{app:conjecture}

\begin{figure}
  \centering
  \includegraphics[width=.8\columnwidth]{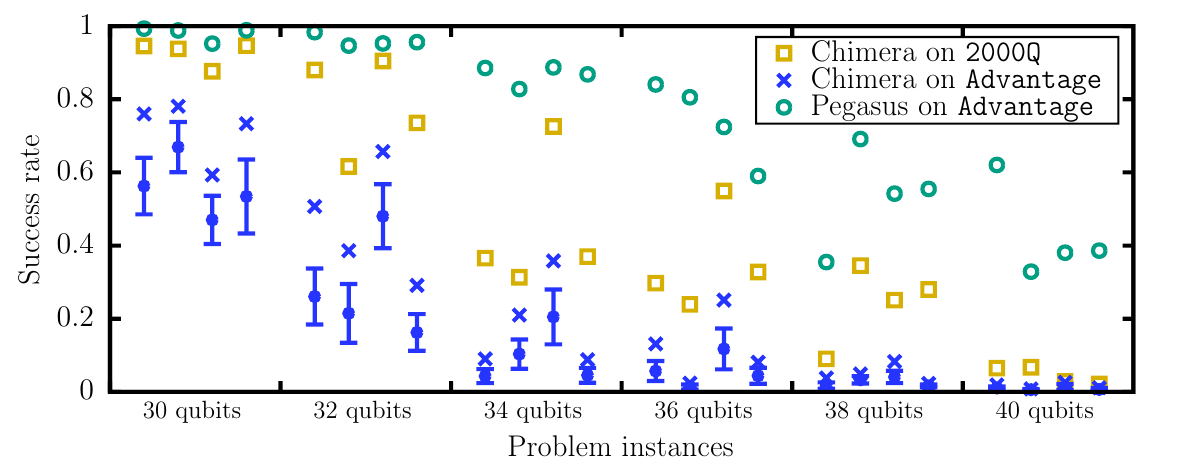}
  \caption{\textbf{Additional experiments to study the conjecture about unused couplers.}
  Shown are the maximum success rates found for each embedding and relative chain strength, using the best Chimera embedding on \TWOOOQ (yellow squares), the same Chimera embedding but used on \Advantage (blue crosses), and the best Pegasus embedding on \Advantage (green circles). Additionally, for ``Chimera on \Advantage'', solid blue circles and error bars indicate the mean and standard deviation, respectively, over all different positions where the Chimera embedding can be placed on the Pegasus graph of \Advantage (i.e., over all possible values of $\Delta x$, $\Delta y$, and $z$ in Eq.~(\ref{eq:chimeraonpegasus})).}
  \label{fig:conjecture}
\end{figure}

Based on the results for the sparse $50$--$120$ qubit problems presented in Fig.~\ref{fig:dwavelargeproblems}, we formulated the conjecture that the additional couplers physically present on \Advantage may disturb the results if they are not needed. One might wonder whether the effect could be mitigated by using logical chains with redundant physical qubits in the embeddings.

To provide additional support for our conjecture from the dense $30$--$40$ qubit problems, we perform the following experiments: We take the best Chimera embedding found for \TWOOOQ (for problem instance 0, these correspond to the maxima in Fig.~\ref{fig:scanrcs}(a)) and use this same embedding to solve the problem on \Advantage. 

Since the Pegasus graph on \Advantage is quite large, it can embed the Chimera subgraphs in several different places (cf.~Fig.~2 in \cite{calaza2021gardenoptimization}). We run experiments for all possible relative displacements $\Delta x\in\{-15,\ldots,15\}$, $\Delta y\in\{-15,\ldots,15\}$, and $z\in\{0,1,2\}$ that generate valid embeddings. 
Physical qubits $q_C\in\{0,\ldots,2048-1\}$ on the Chimera graph are mapped to physical qubits $q_P\in\{0,\ldots,5760-1\}$ on the Pegasus graph using the formula
\begin{align}
    \label{eq:chimeraonpegasus}
    q_P = 60 + 180x + y + 60z + 15(q_C\,\mathrm{mod}\,4) + 
    \begin{cases}
      0 & \text{(if $q_C\,\mathrm{mod}\,8=0,1,2,3$)} \\
      3060 - 179(x-y) - 120z & \text{(if $q_C\,\mathrm{mod}\,8=4,5,6,7$)} \\
    \end{cases},
\end{align}
where $x=\lfloor(q_C\,\mathrm{mod}\,128) / 8\rfloor + \Delta x$, $y=\lfloor q_C / 128\rfloor + \Delta y$, and $z=0,1,2$ denote the corresponding unit cell within the graphs (this formula has been obtained by comparing and relating the different physical qubit labels on \TWOOOQ and \Advantage; the same result can also be achieved using tools from D-Wave NetworkX \cite{DWNetworkX}).

The results are shown in Fig.~\ref{fig:conjecture}.
For each valid embedding, we scan the relative chain strengths (analogously to Fig.~\ref{fig:scanrcs}). We only consider results for the best relative chain strengths found for each embedding.
As expected from Fig.~\ref{fig:scanrcs}, the best Pegasus embeddings on \Advantage (green circles) outperform the best Chimera embeddings on \TWOOOQ (yellow squares). However, the Chimera embeddings on \Advantage (blue crosses) perform significantly worse than on \TWOOOQ. In particular, as the blue circles with error bars show, this worse performance is independent of ($\Delta x$,$\Delta y$,$z$), i.e., independent of where the Chimera embeddings are placed on the larger Pegasus graph.

We draw two conclusions from these results: First, the better overall performance on \Advantage can be tied most strongly to the improved connectivity of the device, which causes the best embeddings to have much shorter chain lengths (cf.~Fig.~\ref{fig:problemdetails}(b)). Second, when the same embeddings (and thus the same chains) are used on both devices, the main difference lies in the many additional unused physical qubits coupling to the chains. Therefore, the significantly worse performance shown in Fig.~\ref{fig:conjecture} provides direct support for the conjecture that the extra couplers on \Advantage can negatively impact the performance if they are not needed.

\section{Exact cover problem details}
\label{app:details}

\begin{figure}
  \centering
  \includegraphics[width=.8\columnwidth]{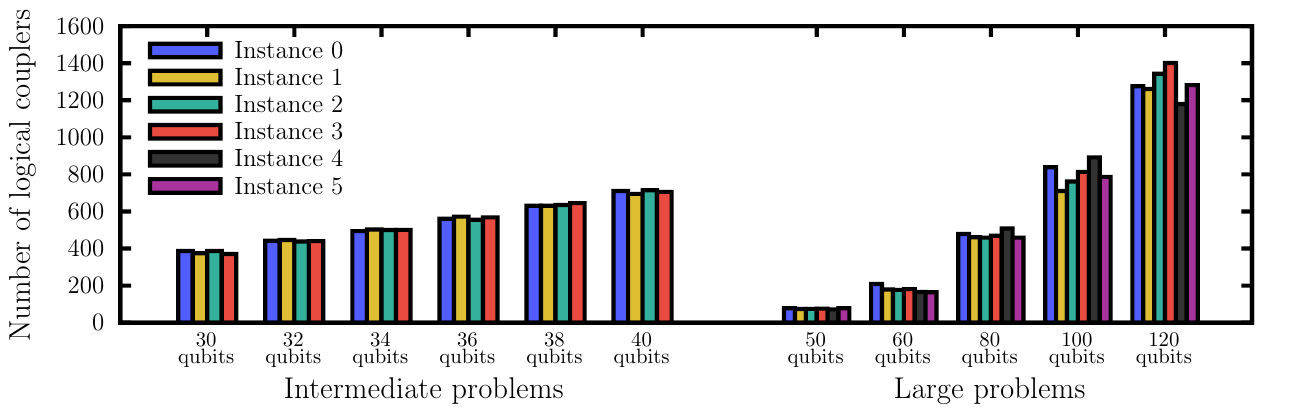}
  \caption{\textbf{Number of logical couplers required for the exact cover problems in the present benchmark set.}
  The intermediate problems with 30--40 qubits contain 4 different problem instances for each problem size (left).
  The large problems with 50--120 qubits contain 6 different problem instances (right).
  Note that after the embedding, the numbers of physical qubits and couplers required on the QPUs
  may be much larger (cf.~Fig.~\ref{fig:problemdetails}(a)).}
  \label{fig:nonzerocouplers}
\end{figure}

\begin{figure}
  \centering
  \includegraphics[width=.5\columnwidth]{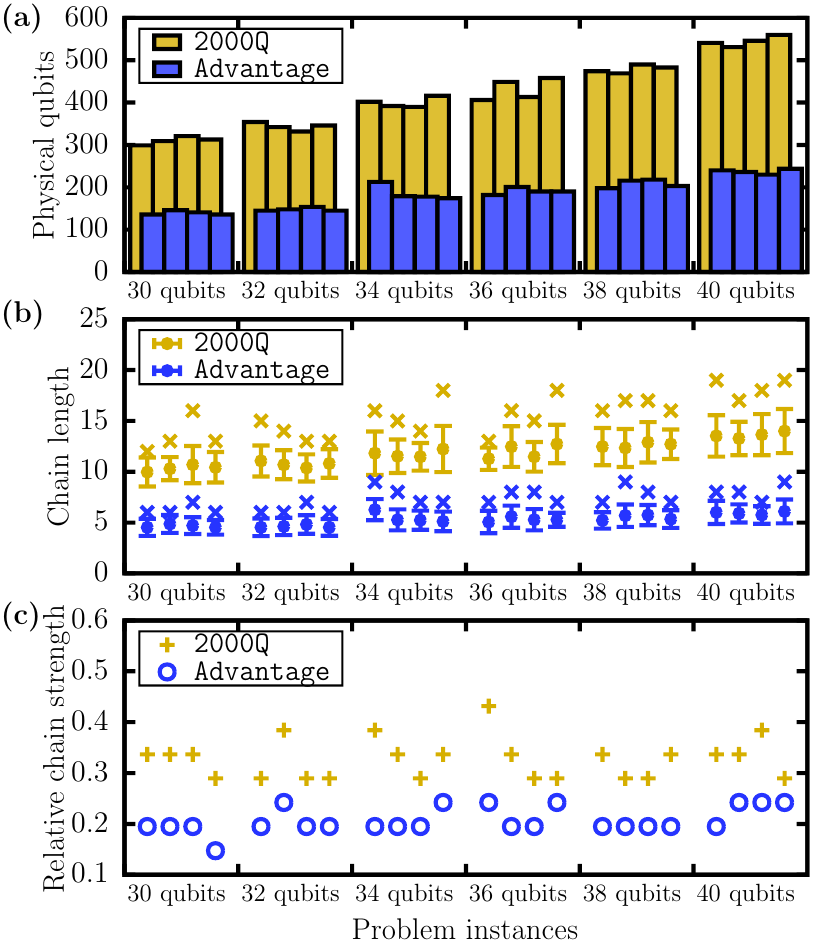}
  \caption{\textbf{Details on the optimal embeddings found for the intermediate exact cover problems
  with $N=30,32,34,36,38,40$ logical qubits on \TWOOOQ (yellow) and \Advantage (blue).}
  For each $N$, the four problem instances
  0, 1, 2, and 3 are shown from left to right on the bottom axes.
  \textbf{(a)} Number of physical qubits needed for the best embeddings;
  \textbf{(b)} lengths of the corresponding physical qubit chains (cf.~Sec.~\ref{sec:qubits}), where the markers with
  error bars indicate the mean and the standard deviation, and crosses indicate
  the maximum chain lengths; \textbf{(c)} optimal relative chain strengths, taken from
  the positions of the corresponding peaks in Fig.~\ref{fig:scanrcs}.}
  \label{fig:problemdetails}
\end{figure}

In this appendix, we provide details on the exact cover problems used in the present benchmark set. For each problem instance, Fig.~\ref{fig:nonzerocouplers} shows the number of couplers required between the qubits. Note that the 30--40 qubit problems require almost all-to-all connectivity.

In Fig.~\ref{fig:problemdetails}, we provide details on the generated embeddings for the intermediate 30--40 qubit problems. In Fig.~\ref{fig:problemdetails}(a),
we list the number of physical qubits required in the embeddings on \TWOOOQ and \Advantage. As these problems require almost full connectivity, the number of physical qubits is much larger than the number of logical qubits, especially on \TWOOOQ. We see that \TWOOOQ needs more than twice as many physical qubits, but also the slope as a function of the logical qubits
is steeper. This is reasonable as also the number of logical couplers increases
(cf.~Fig.~\ref{fig:nonzerocouplers}), and with sparser connectivity on the Chimera topology,
more physical qubits need to be chained into a logical qubit.

This trend is also visible when looking at the chain lengths shown in Fig.~\ref{fig:problemdetails}(b):
While the chains on \Advantage stay almost constant for increasing problem sizes,
the chains on \TWOOOQ grow longer on average. Especially for 40 qubits, chains
on \TWOOOQ can be up to 19 physical qubits long. Such chains are almost always broken
and may lead to a wrong value for the logical qubit that they represent (cf.~Sec.~\ref{sec:qubits}).

In Fig.~\ref{fig:problemdetails}(c), we plot the relative chain strengths that
produced the best results. For each $N=30,32,34,36,38,40$, the first point (instance 0)
corresponds to the peak with the optimal success rate in the corresponding panel
in Fig.~\ref{fig:scanrcs}.

\section{Tail assignment problems with \texorpdfstring{$\lambda\neq0$}{lambda!=0}}
\label{app:lambda}

\begin{figure}
  \centering
  \includegraphics[width=.45\columnwidth]{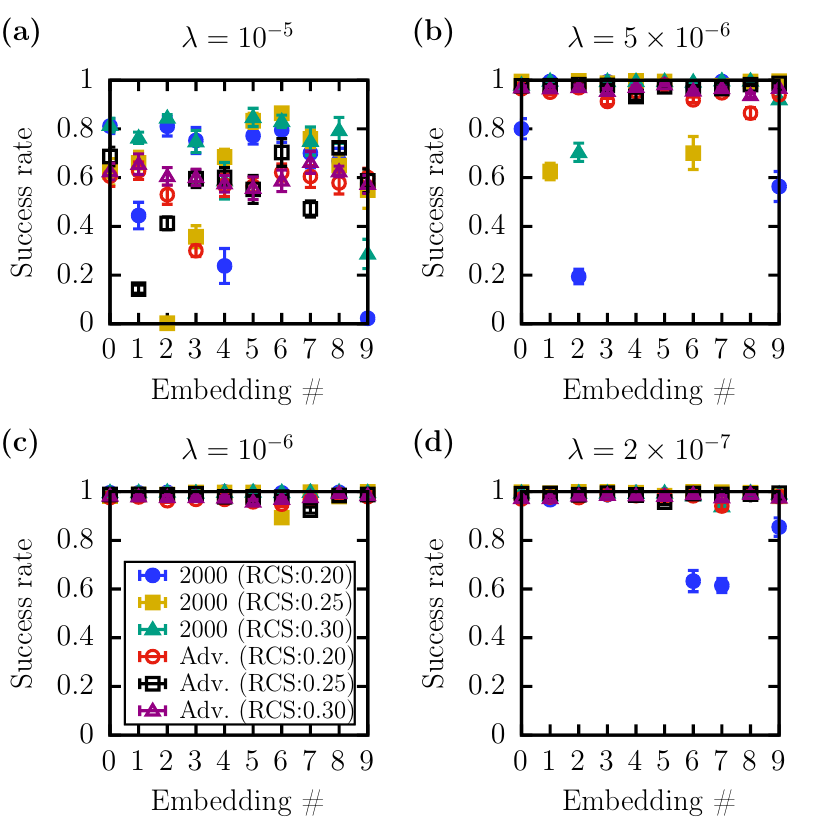}
  \caption{\textbf{Success rates for the tail assignment problem on \TWOOOQ (filled markers) and \Advantage (unfilled markers) as a function of the relative chain strength RCS given by \equref{eq:rcs} (see legend).} Each run is performed for 10 different embeddings (bottom axis). For a given embedding, the markers indicate the mean and the standard deviation for 10 repetitions using the same embedding. The formulation of the tail assignment problem given by \equref{eq:TailAssignmentBQM} for 25 qubits is solved for decreasing values of the scaling factor
  \textbf{(a)} $\lambda=10^{-5}$,
  \textbf{(b)} $\lambda=5\times10^{-6}$,
  \textbf{(c)} $\lambda=10^{-6}$,
  \textbf{(d)} $\lambda=2\times10^{-7}$.}
  \label{fig:TailAssignment}
\end{figure}

The tail assignment problem introduced in Sec.~\ref{sec:tailassignment} contains an objective function that represents the cost associated with each flight route (cf.~\equref{eq:TailAssignmentProblemCost}).
Depending on the magnitude of these cost terms, the multiplier $\lambda$ in the BQM version of the problem (see \equref{eq:TailAssignmentBQM}) has to be adjusted to put a reasonable weight on the objective function with respect to the constraints. Therefore, we test several values of $\lambda$ for a 25 qubit problem. For each $\lambda$, we generate 10 embeddings (cf.~Sec.~\ref{sec:qubits}) and evaluate the success rate on \TWOOOQ and \Advantage. The unique ground state was found using both a linear program solver and exact enumeration of all $2^{25}$ states on a GPU.
We remark that the 25 qubit tail assignment problem solved in this section is the same problem that was investigated as the largest problem instance in \cite{Vikstal2019QAOATailAssignment}.

The results as a function of the embedding are shown in Fig.~\ref{fig:TailAssignment}. We see that for relatively large $\lambda$, the success rates for \TWOOOQ fluctuate strongly as a function of the embedding. In such cases, it may become possible to violate some of the constraints to obtain a better value of the objective function (cf.~\equref{eq:TailAssignmentBQM}). However, as $\lambda$ approaches zero (Figs.~\ref{fig:TailAssignment}(b)--(d)) and the problem approaches its exact cover version, most embeddings yield the optimal solution with unit probability,
especially on \Advantage (unfilled markers).

\end{document}